\documentclass[12pt]{article}
\usepackage{a4wide,epsfig,psfrag,amsmath,amssymb,cite,scalefnt}
\usepackage[dvipsnames]{xcolor}
\usepackage{amsmath,comment,braket}
\usepackage{placeins}
\usepackage{breqn}
\usepackage{slashed}
\usepackage{comment}
\usepackage{subcaption}
\usepackage{hyperref}

\graphicspath{ {figs/} }

\newcommand{\Li}{{\rm Li}}

\usepackage{listings}
\allowdisplaybreaks

\begin{document}

\title{\vskip-3cm{\baselineskip14pt
    \begin{flushleft}
      \normalsize CERN-TH-2025-245, P3H-25-101, TTP25-047
    \end{flushleft}} \vskip1.5cm
  Three-loop corrections to $gg\to ZH$ in the large top quark mass limit
}

\author{
  Joshua Davies$^{a}$,
  Dominik Grau$^{b}$,
  Kay Sch\"onwald$^{c}$,
  \\
  Matthias Steinhauser$^{b}$,
  Daniel Stremmer$^{b}$
  \\[1mm]
  {\small\it (a) Department of Mathematical Sciences, University of
    Liverpool,}
  {\small\it Liverpool, L69 3BX, UK}
  \\
  {\small\it (b) Institut f{\"u}r Theoretische Teilchenphysik}\\
  {\small\it Karlsruhe Institute of Technology (KIT)}\\
  {\small\it Wolfgang-Gaede Stra\ss{}e 1, 76131 Karlsruhe, Germany}
  \\
  {\small\it (c) Physik-Institut, Universit\"at Z\"urich, Winterthurerstrasse 190,}\\
  {\small\it 8057 Z\"urich, Switzerland}
}

\date{}

\maketitle

\thispagestyle{empty}

\begin{abstract}

  We compute three-loop virtual corrections to the associated
  production of a Higgs boson with a $Z$ boson in the large-$m_t$
  limit.  We describe in detail the application of the asymptotic
  expansion and provide, for all form factors,
  analytic results for the first three terms in
  the $1/m_t$ expansion.
  We also provide numerical routines implemented in the \texttt{C++} library \texttt{ggxy}.
  
\end{abstract}


\thispagestyle{empty}

\newpage


\section{Introduction}

The associated production of a Higgs boson with a vector boson,
$W^{\pm}$ or $Z$, provides the third-largest cross section for Higgs
boson production at the LHC. Furthermore, it is an important process
for the measurement of the Higgs coupling to bottom
quarks~\cite{ATLAS:2018kot,CMS:2018nsn,ATLAS:2023jdk}, which is
difficult in other Higgs production processes due to large
backgrounds.

For the quark-initiated channel for $VH$ production, the inclusive
cross section is known up to next-to-next-to-next-to-leading order
(N$^3$LO) in the strong
coupling~\cite{Han:1991ia,Brein:2003wg,Baglio:2022wzu} (see also the
public code \texttt{VH@NNLO}~\cite{Brein:2012ne,Harlander:2018yio}
which includes next-to-next-to-leading order (NNLO) corrections).
Differential cross sections have
been computed in Refs.~\cite{Ferrera:2014lca,Campbell:2016jau} and
electroweak corrections are known from
Refs.~\cite{Ciccolini:2003jy,Denner:2011id}. They are implemented in
the code \texttt{HAWK} \cite{Denner:2014cla}.

In contrast to $WH$ production, $ZH$ production receives
contributions from a gluon-initiated subprocess which has been
computed at leading order (LO) in
Refs.~\cite{Dicus:1988yh,Kniehl:1990iva}. It comes with a
$\mathcal{O}(25)\%$ scale uncertainty which induces a
3\%
uncertainty~\cite{LHCHiggsCrossSectionWorkingGroup:2016ypw} on the
complete $ZH$ contribution. Although the LO $gg\to ZH$ process is a
NNLO contribution to the $pp\to ZH$ cross section, 
higher-order corrections to $gg\to ZH$
are numerically important. This is particularly true in the boosted
regime~\cite{Englert:2013vua} where the $gg$-initiated channel gains
relative importance with respect to the $q\bar{q}$ channel.  In order
to reduce the scale uncertainties in $ZH$ production, next-to-leading order
(NLO) corrections to
$gg\to ZH$ have been computed in several different works.  In
\cite{Altenkamp:2012sx} the infinite-top-mass limit has been applied
and in~\cite{Hasselhuhn:2016rqt,Davies:2020drs} an expansion in
$1/m_t$ has been performed. High-energy results have been obtained in
Refs.~\cite{Davies:2020drs,Davies:2025out} and expansions around the
forward limit are available
from~\cite{Alasfar:2021ppe,Davies:2025out}. Numerical results have
been obtained in Refs.~\cite{Chen:2020gae} and~\cite{Wang:2021rxu},
where in the latter case an expansion in the external masses has been
performed.  There are various approaches where either numerical
results or different expansions are combined in order to cover the
whole phase space; in Refs.~\cite{Chen:2022rua} the numerical results
from~\cite{Chen:2020gae} and the high-energy
expansion~\cite{Davies:2020drs}, in Ref.~\cite{Alasfar:2021ppe} the
forward expansion and the high-energy
expansion~\cite{Davies:2020drs}, and in Ref.~\cite{Davies:2025out}
deep expansions around the forward and high energy limits, see also
Ref.~\cite{CampilloAveleira:2025rbh}.

The aim of this paper is to provide a first step towards NNLO for
$gg\to ZH$ by computing three expansion terms for 
the virtual three-loop corrections, in the large $m_t$
limit. Although the radius of convergence is restricted to 
a relatively small region in phase space, below the $t\bar{t}$
threshold, the large-$m_t$ calculation serves as benchmark for
future exact calculations or calculations in other kinematic limits.

In this paper we provide, in particular, the three-loop virtual corrections 
in the infinite top quark mass limit
which is an important ingredient for the construction
of approximate NNLO predictions for $gg\to ZH$.
An approximation could be constructed in analogy to the
approach outlined in Ref.~\cite{Grazzini:2018bsd} for $gg\to HH$, 
in which exact results
are used up to NLO and for the double-real emission at NNLO,
and virtual corrections at NNLO are approximated in the infinite top quark mass limit,
re-weighted using NLO results. 

The remainder of the paper is structured as follows: in the next
section we provide details of the asymptotic expansion in $1/m_t$
and briefly discuss the various integral families which appear. Results for the 
form factors are presented in Section~\ref{sec::ggZH}.
We conclude in Section~\ref{sec::concl}. In the Appendix we describe the implementation of our results in the \texttt{C++} library \texttt{ggxy}~\cite{Davies:2025qjr}.

\section{Technical details}

\begin{figure}[t]
    \centering
    \includegraphics[trim= 20 590 20 20, width=\textwidth]{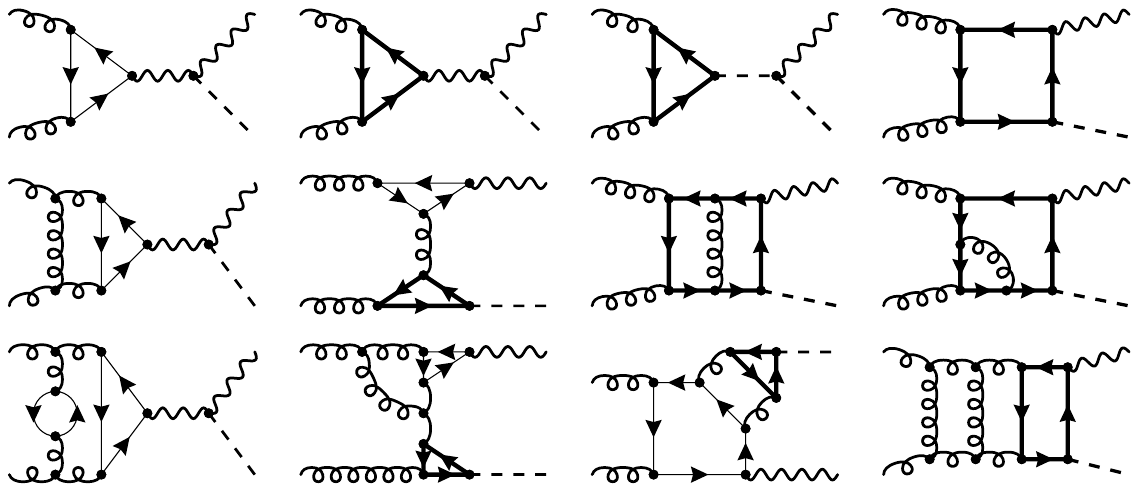}
  \caption{\label{fig::diags}
  One-, two- and three-loop sample Feynman diagrams
  contributing to $gg\to ZH$. Solid thin (thick) lines
  denote massless (massive) quarks. Scalars and gluons are represented by dashed and curly lines, respectively. In the triangle diagrams
  either a $Z$ boson or Goldstone boson mediates the
  coupling to the Higgs and $Z$ boson in the final state.
  At two- and three-loop order both one-particle reducible and one-particle irreducible diagrams have to be considered. These Feynman diagrams were drawn with the help of the \textsc{FeynGame} program \cite{Bundgen:2025utt}.}
\end{figure}

We consider the scattering of two gluons in the initial state with
momenta $q_1$ and $q_2$ into a Higgs and a $Z$ boson 
with momenta $q_3$ and $q_4$. The Mandelstam variables are
then given by
\begin{eqnarray}
  s = (q_1+q_2)^2\,,\qquad t = (q_1+q_3)^2\,,\qquad u = (q_1+q_4)^2\,,
  \label{eq::stu}
\end{eqnarray}
where all momenta are incoming. Furthermore we have
\begin{eqnarray}
  q_1^2=q_2^2=0\,\qquad
  q_3^2=m_Z^2\,,\qquad q_4^2=m_H^2\,.
  \label{eq::qs}
\end{eqnarray}
The transverse momentum of the final-state particles is given by
\begin{eqnarray}
  p_T^2 &=&\frac{u\,t-q_3^2 q_4^2}{s}\,.
            \label{eq::pT}
\end{eqnarray}
The (internal) top quark mass is denoted by $m_t$.
In Fig.~\ref{fig::diags} we show sample Feynman diagrams 
for the virtual corrections to $gg\to ZH$ up to three-loop order (NNLO).

We decompose the amplitude for $g(q_1)g(q_2) \to Z(q_3)H(q_4)$ as a
linear combination of form factors following
Ref.~\cite{Kniehl:1990iva}; we use the notation of
Refs.~\cite{Davies:2020drs,Davies:2025out} and obtain six independent
form factors given by
\begin{align} \label{eq:ffs}
    F_{12}^+(t,u),\: F_{12}^-(t,u),\: F_2^-(t,u),\: F_3^+(t,u),\: F_3^-(t,u),\: F_4(t,u)
    \,,
\end{align}
see Eqs.~(34) and (36) of Ref.~\cite{Davies:2025out} for more details. 

In the computation of $gg \to ZH$ we have to take into account the contributions from light and heavy quarks inside the loops. The contributions proportional to the vector coupling vanish exactly. 
Therefore, only the axial-vector coupling has to be taken into account. Since the axial-vector coupling is proportional to the third component of the weak isospin of the quark $I^3_{u,d}=\pm 1/2$, the contributions of quark generations with equal masses vanishes. Considering the top quark as the only massive quark, we have to take into account only the axial-vector coupling of the $Z$ boson to top and bottom quarks. 

The amplitude is generated with \texttt{qgraf}~\cite{Nogueira:1991ex} where we encounter $23$ diagrams at the one-loop, $398$ at the two-loop and $11,866$ diagrams at the three-loop level.\footnote{Goldstone bosons are taken into account and the longitudinal mode of the $Z$ boson is treated as an separate particle.}
The output is converted to \texttt{FORM}~\cite{Ruijl:2017dtg} code with the combination of the tools \texttt{q2e} and
\texttt{exp}~\cite{Harlander:1998cmq,Seidensticker:1999bb}, where the latter tool also performs the mapping to topologies including subgraphs and co-subgraphs as required by the large mass expansion and described in more detail below. Further calculation is then performed with the in-house setup \texttt{calc}, where the calculation of the Dirac and colour traces is performed and the amplitude is expanded in the limit of the large top-quark mass. The result is then given as a sum of scalar integrals for which the integration-by-parts (IBP) reduction and insertion of master integrals has to be still carried out.


\subsection{Asymptotic expansion}

\begin{figure}[t]
    \centering
    \includegraphics[trim= 20 625 20 20, width=\textwidth]{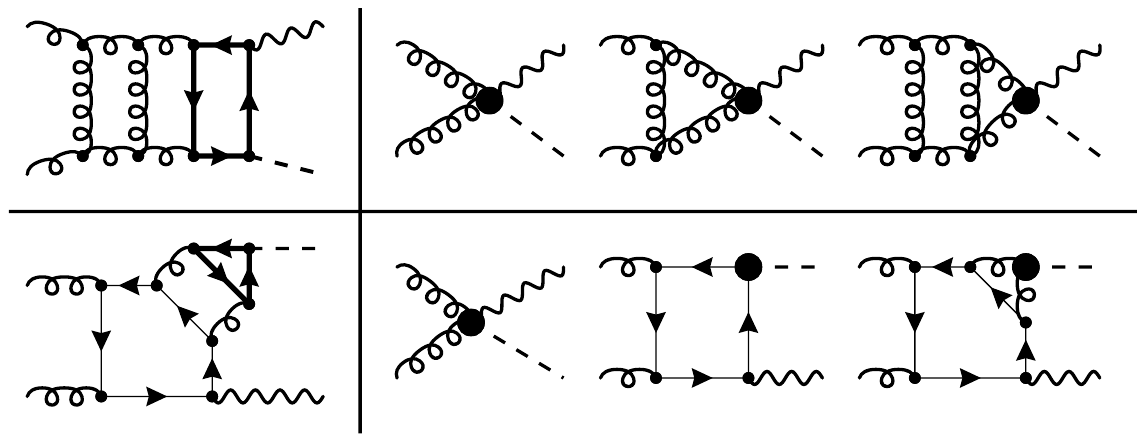}
  \caption{\label{fig::subgraphs}
Graphical representation of the asymptotic expansion applied
to the diagrams on the left. On the right we show only the
co-subgraphs. The corresponding subgraphs are one-, two- and three-loop vacuum integrals which are inserted in the
effective vertices represented by the blobs. 
}
\end{figure}

We use \texttt{exp}~\cite{Harlander:1998cmq,Seidensticker:1999bb} to apply the hard mass expansion procedure (see, e.g., Ref.~\cite{Smirnov:2002pj})
to the three-loop triangle and box diagrams. 
This leads to a number of subgraphs and co-subgraphs;
see Fig.~\ref{fig::subgraphs} for typical examples in
graphical from.

In our calculation the following cases appear:
\begin{itemize}
\item Three-loop massive subgraph. Here the co-subgraph is
  a tree-level diagram. For the three-loop vacuum graphs we need
  tensor integrals up to rank $10$.
\item In case the subgraph is a two-loop vacuum integral
  the co-subgraph is a massless integral with up to four external
  legs, two of which can be massive ($m_H$ and $m_Z$). 
  The corresponding integrals are well-known in the literature
  (see, e.g., Ref.~\cite{Ellis:2007qk}).
\item The most complicated case is the one where the 
  subgraph is a one-loop vacuum graph. Here the 
  co-subgraph is in general a massless double box integral
  with two massless legs and two massive legs with masses
  $m_H$ and $m_Z$. Below we provide
  more details about the calculation of this contribution.
\item
  Note that the purely massless diagrams which involve a closed
  light-quark loop are computed exactly.
  These are triangle diagrams, where massless form factor integrals are needed. 
\end{itemize}

The main differences to the computation of the large mass expansion of $gg\to HH$~\cite{Davies:2019djw} originate
from the latter two points, see also diagrams as shown in Fig.~\ref{fig::diags}. Integral families for two-loop massless box integrals and three-loop massless vertices do not appear for Higgs boson pair production.

One-loop massless box integrals, vacuum integrals up to three loops
and massless triangle integral families up to three
loops are built into the \texttt{calc} setup. This includes also routines
to perform the tensor decomposition.
Additional work is necessary for the two-loop massless box topologies. In 
that case we use \texttt{tapir}~\cite{Gerlach:2022qnc} to generate two-
loop massless box families onto which we map our scalar integrals. The IBP 
reduction is performed with \texttt{Kira}~\cite{Klappert:2020nbg}, where 
we reduce the scalar integrals to a set of master integrals. These are 
members of the two-loop massless box families 
introduced in Ref.~\cite{Gehrmann:2015ora}.
In a second step, these master 
integrals are rewritten as linear combinations of the canonical basis 
defined in Refs.~\cite{Henn:2014lfa,Caola:2014lpa}, where they have also 
been calculated for the first time. We use the analytic 
results presented in Ref.~\cite{Gehrmann:2015ora}, 
consisting of $84$ master integrals, plus crossings of external particles, 
in terms of a so-called ``optimized functional basis'' including logarithms, 
polylogarithms $\Li_n$ ($n=2,3,4$) and $\Li_{2,2}$ functions. We note 
that the master integrals are only needed up to weight $3$, so that the 
functions $\Li_4$ and $\Li_{2,2}$ are not present in our final results. In addition, we perform a multivariate partial fractioning with the tool \texttt{MultivariateApart}~\cite{Heller:2021qkz} to simplify the complicated rational functions arising from these topologies. This reduces the size of the final expressions by a factor of $10$ to $15$.

\subsection{UV renormalization and IR subtraction}

The UV renormalization and IR subtraction closely follows Ref.~\cite{Davies:2019djw}. In particular, we renormalize the top-quark mass and the gluon wave function in the on-shell scheme and the strong coupling constant $\alpha_s^{(6)}$ in the $\overline{\rm MS}$ scheme. In addition, we also need the decoupling constant of $\alpha_s$ to write the final results in terms of $\alpha_s^{(5)}$. A summary of all expressions can be found e.g.~in Ref.~\cite{Davies:2025ghl}. In the process $gg\to ZH$, $\gamma_5$ 
is present in the axial-vector coupling of the $Z$ boson and in the pseudo-scalar coupling of the Goldstone boson. We adapt the prescription from Ref.~\cite{Larin:1993tq}.
Thus, we have to renormalize the non-singlet axial-vector and pseudo-scalar currents with $Z^{ns}= Z^{ns}_fZ^{ns}_{\overline{\rm{MS}}}$ and $Z^{ps}= Z^{ps}_fZ^{ps}_{\overline{\rm{MS}}}$, respectively. The corresponding renormalization constants are given by \cite{Larin:1993tq,Chen:2024cvu} 
\begin{eqnarray*}
  Z^{ns}_f &=& 1 - 4a_s C_F +a_s^2\left(22 C_F^2 - \frac{107}{9} C_A C_F + \frac{4}{9} C_F T_F n_f\right)\,,\nonumber\\
  Z^{ps}_f &=& 1 - 8 a_s C_F + a_s^2\left(\frac{2}{9} C_A C_F + \frac{8}{9} C_F T_F n_f\right)\,,
\end{eqnarray*}
and
\begin{eqnarray*}
  Z^{ns}_{\overline{\rm{MS}}} &=& 1 +a_s^2\left(\frac{22}{3\epsilon} C_A C_F - \frac{8}{3\epsilon} C_F T_F n_f\right)\,,\nonumber\\
  Z^{ps}_{\overline{\rm{MS}}} &=& 1 + a_s^2\left(\frac{44}{3\epsilon}C_A C_F - \frac{16}{3\epsilon} C_F T_F n_f\right)\,,
\end{eqnarray*}
with $C_A=3$, $C_F=4/3$, $T_F=1/2$, $a_s=\alpha_s^{(6)}/(4\pi)$ and $n_f=n_h+n_l$, where $n_h=1$ is the number of heavy quarks 
and $n_l=5$ is the number of light quarks.

For the subtraction of IR singularities we follow Refs.~\cite{Catani:1998bh,deFlorian:2012za}, so that the finite form factors are given by
\begin{align}
    \notag F^{(1),{\rm fin}} &= F^{(1)}-\frac{1}{2}I_g^{(1)} F^{(0)},\\
    F^{(2),{\rm fin}} &= F^{(2)}-\frac{1}{2}I_g^{(1)} F^{(1)}-\frac{1}{4}I_g^{(2)} F^{(0)},
\end{align}
where $I_g^{(1)}$ and $I_g^{(2)}$ are given e.g.~in Eqs.~(20) to~(24) in Ref.~\cite{Davies:2019djw} (see also Ref.~\cite{deFlorian:2012za}).\footnote{Note that there is a typo in Eq.~(24) of Ref.~\cite{Davies:2019djw}
and Eq.~(13) or Ref.~\cite{Davies:2025ghl}: In both cases the
sign in front of the $C_An_l$ colour structure should be 
a ($-$) and not a ($+$) sign.}
$F^{(i)}$ are the coefficients of the renormalized 
form factors expanded in $\alpha_s^{(5)}$ as
\begin{equation}
    F=F^{(0)}+\frac{\alpha_s^{(5)}(\mu)}{\pi}F^{(1)}
    +\left(\frac{\alpha_s^{(5)}(\mu) }{\pi}\right)^2 F^{(2)}\,.
    \label{eq::Fi}
\end{equation}


\section{\label{sec::ggZH}Renormalized virtual contribution}

All form factors have been expanded up to ${\cal O}(1/m_t^4)$. We refrain from presenting explicit results for 
the individual form factors in this paper; our
analytic ultra-violet renormalized
results are available in a computer-readable form
from Ref.~\cite{progdata}. Furthermore, they are implemented in \texttt{ggxy}, see Appendix.
We provide both infra-red divergent and infra-red finite expressions.  At one- and two-loop order we reproduce the results in the literature~\cite{Davies:2020drs}.
\begin{table}[t!]
\begin{center}
\begin{tabular}{|c||c|c|c|c|}
\hline 
 & $T_F^3$ & $T_F^2 C_A$ & $T_F^2 C_F$ & $T_F\{C_A^2,C_A C_F,C_F^2\}$ \\ 
\hline 
$F_{12}^+$ & $\mathcal{O}(1/m_t^0)$ & $\mathcal{O}(1/m_t^0)$ & $\mathcal{O}(1/m_t^0)$ & $\mathcal{O}(1/m_t^0)$\\
$F_{12}^-$ & $\mathcal{O}(1/m_t^0)$ & $\mathcal{O}(1/m_t^0)$ & $\mathcal{O}(1/m_t^0)$ & $\mathcal{O}(1/m_t^4)$\\
$F_{2}^-$ & $\mathcal{O}(1/m_t^0)$ & $\mathcal{O}(1/m_t^0)$ & $\mathcal{O}(1/m_t^0)$ & $\mathcal{O}(1/m_t^4)$\\
$F_{3}^+$ & $-$ & $\mathcal{O}(1/m_t^0)$ & $\mathcal{O}(1/m_t^2)$ & $\mathcal{O}(1/m_t^4)$\\
$F_{3}^-$ & $-$ & $\mathcal{O}(1/m_t^0)$ & $\mathcal{O}(1/m_t^2)$ & $\star$\\
$F_{4}$ & $-$ & $\mathcal{O}(1/m_t^0)$ & $\mathcal{O}(1/m_t^2)$ & $\mathcal{O}(1/m_t^4)$\\
\hline 
\end{tabular}
\end{center}
\caption{The table entries show the lowest order of the large-mass expansion at which the different colour factors lead to non-zero contributions to the bare three-loop amplitude. The symbol ``$-$'' denotes that the colour factor vanishes exactly, while ``$\star$'' denotes that it is expected that the colour factor appears in higher-order expansion terms.}
\label{tab:ffcol}
\end{table}

 At one- and two-loop order several form factors start at order $1/m_t^4$. In particular, at one loop the form factor $F_{12}^+$ is the only one which starts at ${\cal O}(1/m_t^0)$, while all other form factors begin at ${\cal O}(1/m_t^4)$. $F_3^-$ even vanishes completely. This situation differs only slightly at the two-loop level. The form factors $F_{12}^-$ and $F_2^-$ now also start at ${\cal O}(1/m_t^0)$ due to the one-particle reducible contributions (double-triangle diagrams), while the one-particle irreducible contributions still vanish at ${\cal O}(1/m_t^0)$ and ${\cal O}(1/m_t^2)$. Additionally, the form factor $F_3^-$ no longer vanishes exactly but begins at ${\cal O}(1/m_t^6)$. On the other hand, at the three-loop level all six form factors have non-zero contributions at ${\cal O}(1/m_t^0)$, while several colour factors first appear at ${\cal O}(1/m_t^2)$ or ${\cal O}(1/m_t^4)$.
 An overview of the order in the large-mass expansion at which the different colour structures first contribute to the bare amplitude is given, for each of the six form factors, in Table \ref{tab:ffcol}.

In the following we provide results for the squared
amplitude which we define at LO, NLO and NNLO
as follows
\begin{eqnarray} \label{eq:A2}
  |A^{(0)}|^2
  \!\!&\!=\!&\!\!\frac{G_F^2 m_Z^2}{16s^2} \!
      \sum_{\lambda_1,\lambda_2,\lambda_3}
      \Bigg\{
      \left[ \tilde{A}_{\rm sub}^{(0),\mu\nu\rho} \tilde{A}_{\rm
      sub}^{(0),\star,\mu^\prime\nu^\prime\rho^\prime} \right]\Bigg\} 
      \nonumber\\&&\mbox{}\qquad \times 
      \varepsilon_{\lambda_1,\mu}(q_1)\:
      \varepsilon^\star_{\lambda_1,\mu^\prime}(q_1)\:
      \varepsilon_{\lambda_2,\nu}(q_2)\:
      \varepsilon^\star_{\lambda_2,\nu^\prime}(q_2)\:
      \varepsilon_{\lambda_3,\rho}(q_3)\:
      \varepsilon^\star_{\lambda_3,\rho^\prime}(q_3)\,,
\nonumber\\
  |A^{(1)}|^2
  \!\!&\!=\!&\!\!\frac{G_F^2 m_Z^2}{16s^2} \!
      \sum_{\lambda_1,\lambda_2,\lambda_3}
      \Bigg\{
      2{\rm Re}\left[\tilde{A}_{\rm sub}^{(0),\mu\nu\rho} \tilde{A}_{\rm
      sub}^{(1)\star,\mu^\prime\nu^\prime\rho^\prime} \right]\Bigg\} 
      \nonumber\\&&\mbox{}\qquad \times 
      \varepsilon_{\lambda_1,\mu}(q_1)\:
      \varepsilon^\star_{\lambda_1,\mu^\prime}(q_1)\:
      \varepsilon_{\lambda_2,\nu}(q_2)\:
      \varepsilon^\star_{\lambda_2,\nu^\prime}(q_2)\:
      \varepsilon_{\lambda_3,\rho}(q_3)\:
      \varepsilon^\star_{\lambda_3,\rho^\prime}(q_3)\,,
\nonumber\\
  |A^{(2)}|^2
  \!\!&\!=\!&\!\!\frac{G_F^2 m_Z^2}{16s^2} \!
      \sum_{\lambda_1,\lambda_2,\lambda_3}
      \Bigg\{
      2{\rm Re}\left[\tilde{A}_{\rm sub}^{(0),\mu\nu\rho} \tilde{A}_{\rm
      sub}^{(2),\star,\mu^\prime\nu^\prime\rho^\prime} \right]
      +\left[ \tilde{A}_{\rm sub}^{(1),\mu\nu\rho} \tilde{A}_{\rm
      sub}^{(1),\star,\mu^\prime\nu^\prime\rho^\prime} \right]\Bigg\} 
      \nonumber\\&&\mbox{}\qquad \times 
      \varepsilon_{\lambda_1,\mu}(q_1)\:
      \varepsilon^\star_{\lambda_1,\mu^\prime}(q_1)\:
      \varepsilon_{\lambda_2,\nu}(q_2)\:
      \varepsilon^\star_{\lambda_2,\nu^\prime}(q_2)\:
      \varepsilon_{\lambda_3,\rho}(q_3)\:
      \varepsilon^\star_{\lambda_3,\rho^\prime}(q_3)\,,
\end{eqnarray}
where $|A|^2$ has the same perturbative expansion as the form factors in Eq.~(\ref{eq::Fi}).
$\tilde{A}_{\rm sub}^{(i)}$
are the infrared-subtracted finite
form factors evaluated for $\mu^2=-s$, see also Ref.~\cite{Davies:2025ghl}.
\begin{figure}[t!]
    \begin{center}
    \includegraphics[width=.45\textwidth]{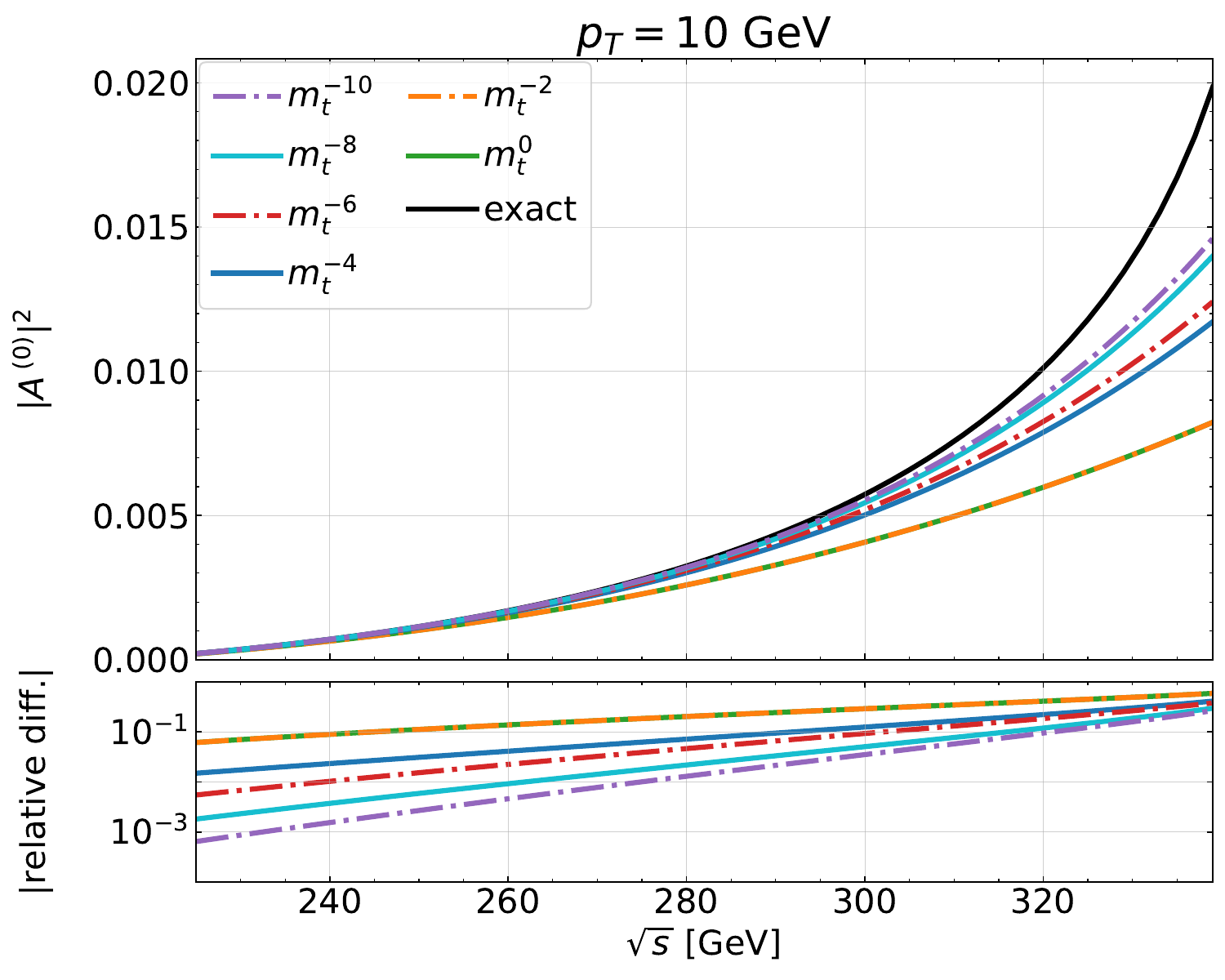}
    \includegraphics[width=.45\textwidth]{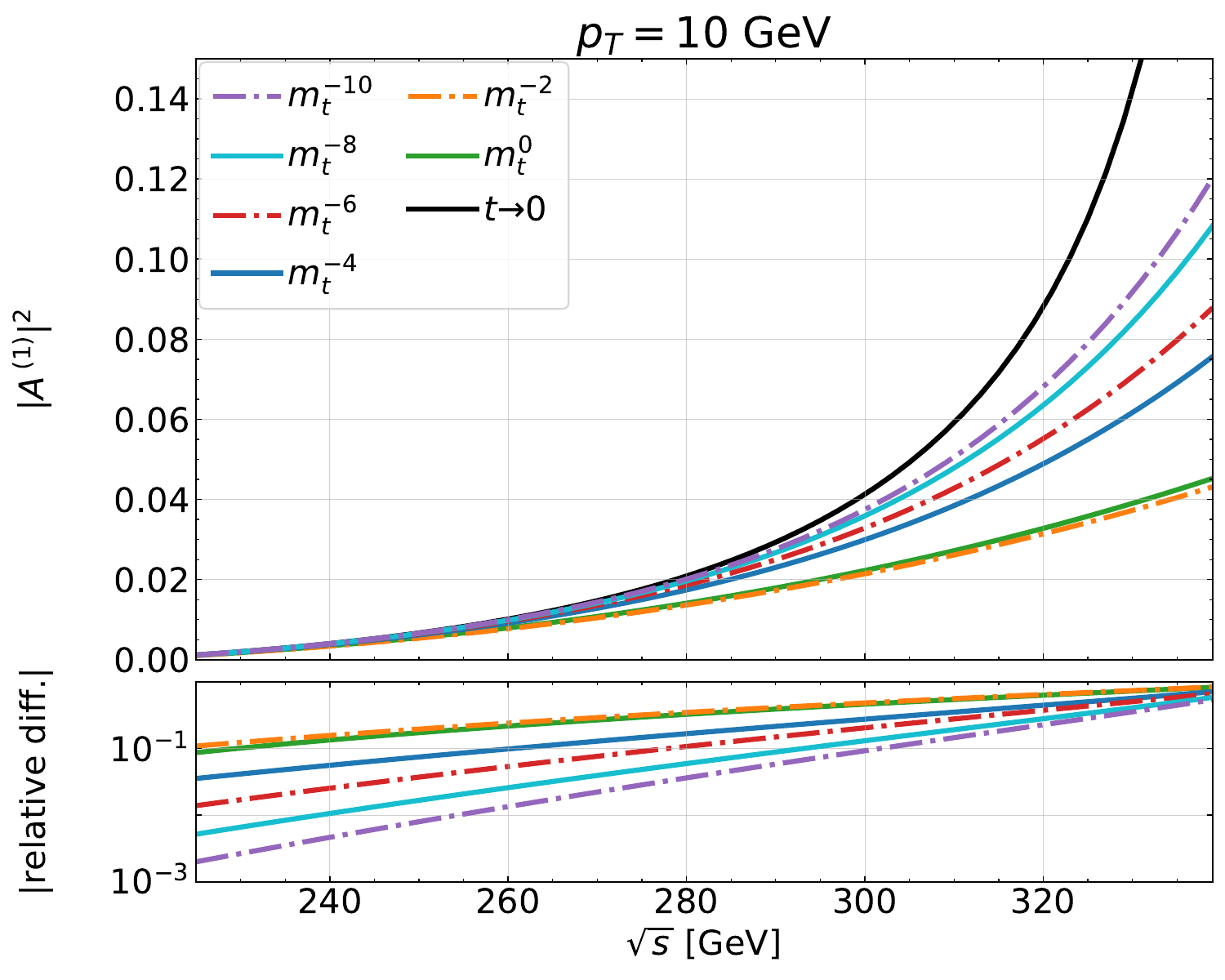}
    \end{center}
  \caption{\label{fig::matsq_pt10}
  Squared amplitude at LO and NLO as a function
  of $\sqrt{s}$ for $p_T=10$~GeV, showing different expansion
  depths in $1/m_t^2$. Exact one-loop results and the results from the forward limit~\cite{Davies:2025out} at two loops are shown in black as a reference. The lower panel displays the relative difference to the reference values.}
\end{figure}
In Fig.~\ref{fig::matsq_pt10} we show
results for $|A^{(0)}|^2$ and
and $|A^{(1)}|^2$ as a function of $\sqrt{s}$ for fixed transverse momentum $p_T=10$~GeV.
In the upper panel we include, step-by-step, higher-order terms in $1/m_t^2$ up to
$\mathcal{O}(1/m_t^{10})$. In addition, we display as reference values the exact results at one-loop order and the results from the expansion in the forward limit~\cite{Davies:2025out} at two loops, which approximates the exact result far below the percent level. In the lower panel the relative difference with respect to the reference results is shown.  

A convergent behaviour of the large-$m_t$ expansion
can be expected only for $\sqrt{s}$ values below $2m_t$.
This is indeed observed at LO. The $1/m_t^2$ terms vanish 
and we observe a relatively large jump once the $1/m_t^4$ terms are included.
For $\sqrt{s}\lesssim 300$~GeV the contribution of higher $1/m_t^2$
terms is numerically less important.
A similar pattern is observed at NLO, where
again the $1/m_t^4$ terms provide a numerically
large contribution.\footnote{Both at LO and NLO we observe a pairing of an even and the subsequent odd expansion term in $1/m_t^2$; this has also been observed for $gg\to HH$ in Ref.~\cite{Grigo:2013rya}.}
Up to $\sqrt{s}\approx 300$~GeV the approximations including $1/m_t^4$ terms
agree at the 10\% level or better with the exact results.\footnote{We remark that after including $1/m_t^{10}$ terms at NLO, the difference reduces to about 3\% for $\sqrt{s}=300$~GeV.}
\begin{figure}[t!]
    \begin{center}
    \includegraphics[width=.45\textwidth]{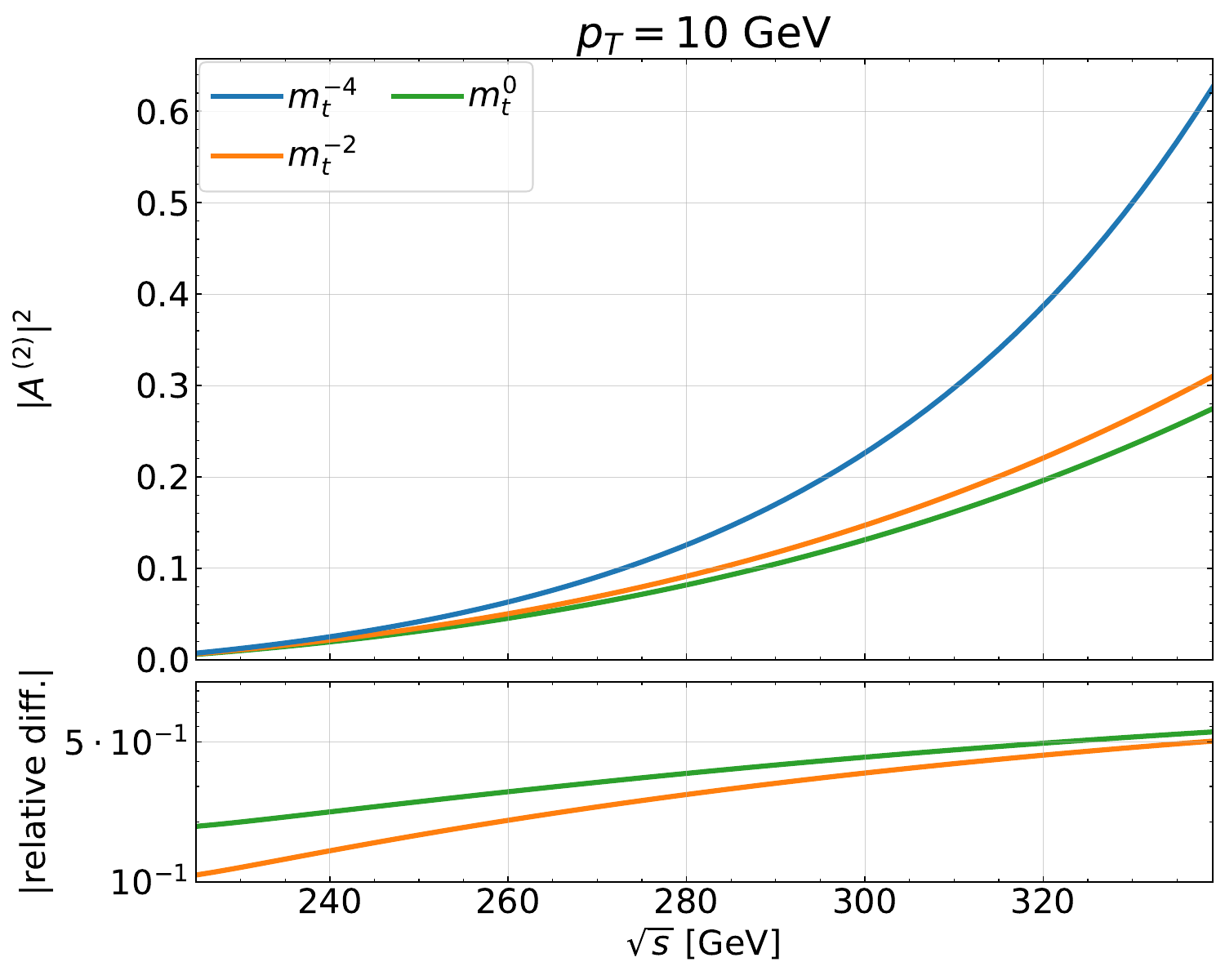}
    \end{center}
  \caption{\label{fig::matsq_pt10_3l}
  Squared amplitude at NNLO as a function
  of $\sqrt{s}$ for $p_T=10$~GeV, showing different expansion
  depths in $1/m_t^2$. In the lower panel the 
  relative difference to the best available approximation ($1/m_t^4$) is shown.}
\end{figure}
We assume that this pattern of convergence extends to NNLO, which is our 
motivation to perform an expansion up to $1/m_t^4$. The computation 
of the next term in the $1/m_t$ expansion would be rather
CPU-time expensive. The numerical results at NNLO up to $1/m_t^4$ are shown in Fig. \ref{fig::matsq_pt10_3l}, where we again find that
the $\mathcal{O}(1/m_t^4)$ terms are numerically important.

\begin{figure}[t!]
    \begin{center}
    \includegraphics[width=.45\textwidth]{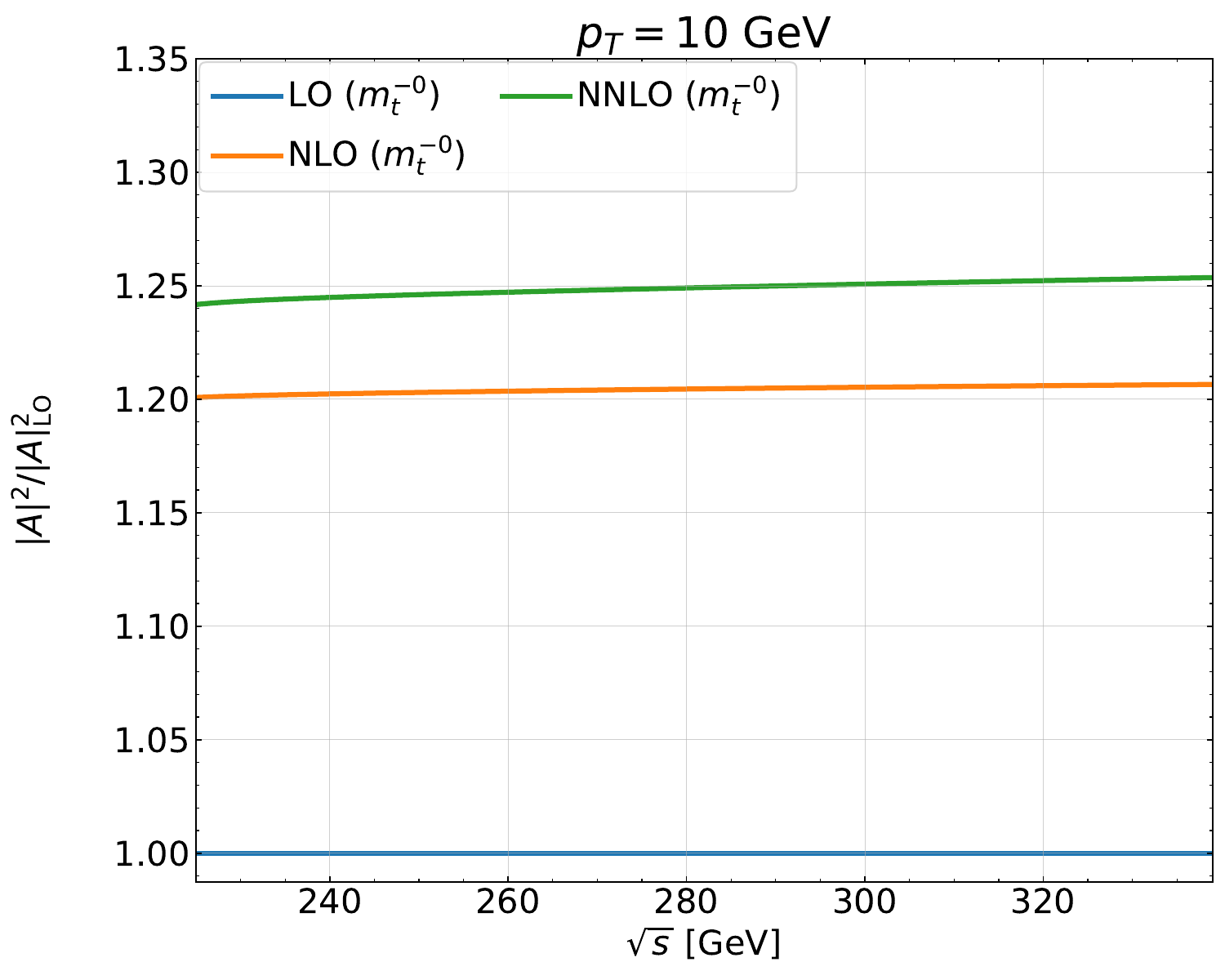}
    \includegraphics[width=.45\textwidth]{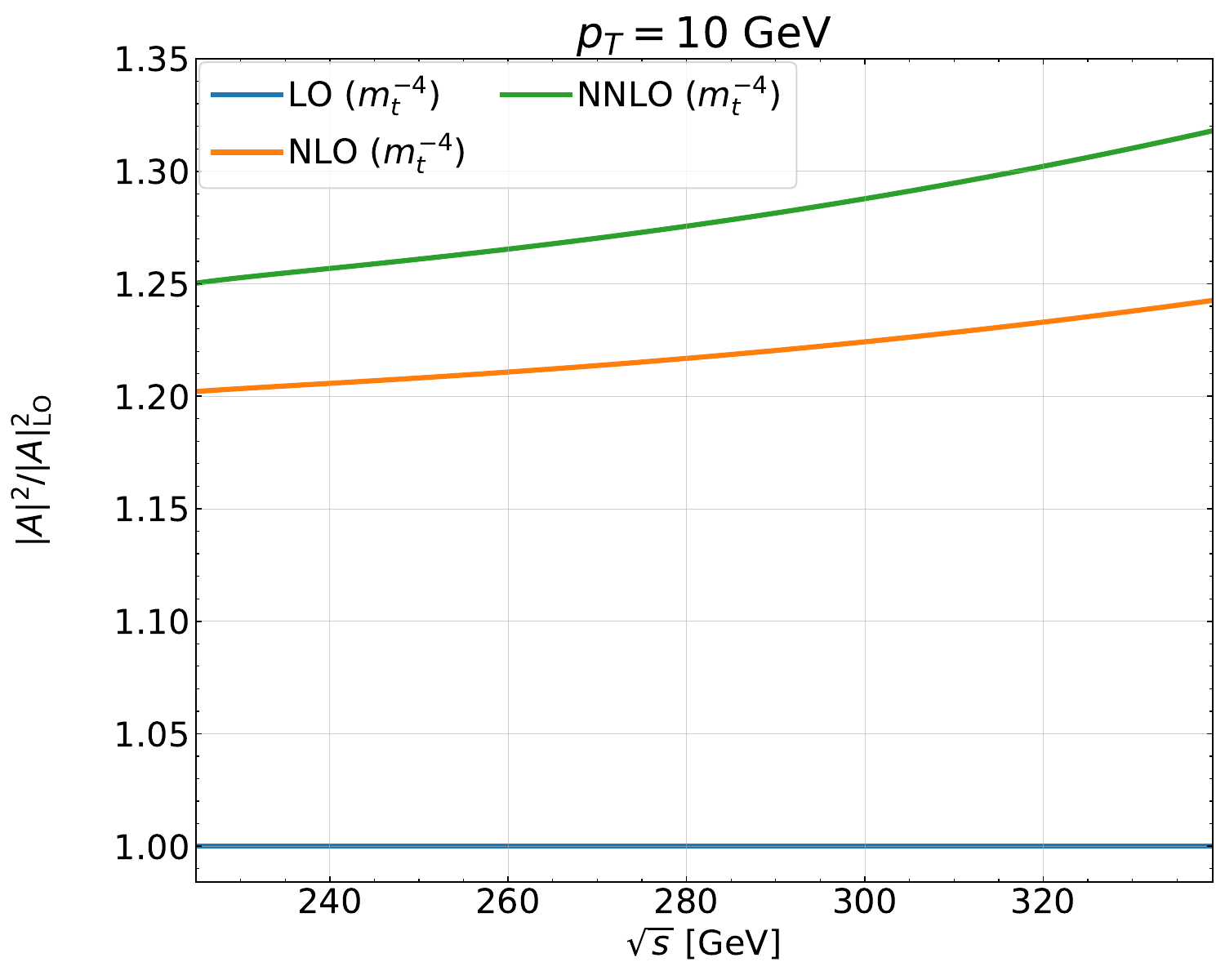}
    \end{center}
  \caption{\label{fig::matsq_pt10_sum}
  Squared amplitude at LO, NLO and NNLO as a function
  of $\sqrt{s}$ for $p_T=10$~GeV normalized to the LO one. Left plot shows the results at the order $1/m_t^0$ and the right plot up to $1/m_t^4$.}
\end{figure}

In Fig.~\ref{fig::matsq_pt10_sum} we show LO, NLO and NNLO results
for the squared amplitude, where we use $\alpha_s=0.118$. In the left panel, only
the infinite top quark mass results are included.
We observe $K$ factors at NLO and NNLO of 20\% and 25\%, respectively, 
which are almost independent of $\sqrt{s}$. 
In the right panel expansions up to $1/m_t^4$ are included
in the calculation of the form factors. Now we observe 
a slight increase on the $K$ factors; at $\sqrt{s}=300$~GeV
they amount to about $22\%$ and $28\%$.


\section{\label{sec::concl}Conclusions}

We have computed the three-loop virtual corrections
to $gg\to ZH$ in the large-$m_t$ limit, taking into account triangle and box contributions. Furthermore, all one-particle irreducible and one-particle reducible diagrams are considered.
This is necessary to obtain a finite result.
We provide an expansion up to $1/m_t^4$. Since some of the form factors only start to contribute at $1/m_t^4$
we observe a relatively big contribution from this
last expansion term. 
The leading term of our expansion provides an
important ingredient for the construction
of an approximate NNLO prediction in analogy to
Ref.~\cite{Grazzini:2018bsd} which is currently
used for $gg\to HH$.
The remaining terms constitute a benchmark result for future numerical calculations or expansions around other limits. Analytic results for all form factors
can be found in Ref.~\cite{progdata}.
They are also implemented in the \texttt{C++} library \texttt{ggxy}~\cite{Davies:2025qjr},
which allows for a fast and convenient numerical evaluation. In particular,
it is straightforward to generate the data for all plots shown in this paper.


\section*{Acknowledgements}  

This research was supported by the Deutsche Forschungsgemeinschaft (DFG,
German Research Foundation) under grant 396021762 --- TRR 257 ``Particle
Physics Phenomenology after the Higgs Discovery''.
The work of J.~D.~was supported by STFC Consolidated Grant ST/X000699/1.
The research of K.~S.~was
funded by the European Union’s Horizon Research and Innovation Program under the Marie Skłodowska-Curie grant agreement No.~101204018.

\begin{appendix}

\section*{Appendix: Implementation in \texttt{ggxy}}

All six form factors of Eq.~\eqref{eq:ffs} are implemented in \texttt{ggxy}, which can be obtained from
\begin{equation*}
\textrm{\url{https://gitlab.com/ggxy/ggxy-release}}.
\end{equation*}
In particular, we include terms up to $m_t^{-10}$ at one- and two-loop and terms up to $m_t^{-4}$ at three-loop order. The finite part of the form factors at the different loop-orders can be evaluated using the functions in the header files \texttt{ff/ggzh/LMEggzh\{1,2,3\}lFF.h},
\begin{lstlisting}
complex<double> LMEggzh{1,2,3}l<FF>(double s, double t,
                                    double mzs, double mhs,
                                    double mts, double mus,
                                    bool irsubtr = true,
                                    unsigned ExpDepth = {5,5,2});
\end{lstlisting}
or
\begin{lstlisting}
vector<complex<double>> LMEggzh{1,2,3}l<FF>imts(double s, double t,
                                            double mzs, double mhs,
                                            double mts, double mus,
                                            bool irsubtr = true);
\end{lstlisting}
where \texttt{<FF>} is a placeholder for the six form factors which are abbreviated as \texttt{FF12p}, \texttt{FF12m}, \texttt{FF2m}, \texttt{FF3p}, \texttt{FF3m} and \texttt{FF4}. The return value is either the form factor up to the order \texttt{ExpDepth} in the $1/m_t^2$ expansion, where the default values correspond to highest number of implemented expansion terms, or a vector containing the coefficients of the expansion. The parameter \texttt{mus} defines the squared renormalization scale ($\mu^2$) and \texttt{irsubtr} is used to (de-)activate the IR subtraction of the two- and three-loop form factors. An example for the numerical evaluation of the form factors is given in the file \texttt{ff-lme.cpp} which can be found in the path \texttt{examples/ggzh-FF}. The example program evaluates the form factors for the phase-space region shown in the main text, where a typical agreement of six digits is found with respect to the evaluation with Mathematica.

Alternatively, the squared amplitudes of Eq. \eqref{eq:A2} can also be directly computed with the function
\begin{lstlisting}
vector<double> ggzh3lAsq(double s, double t, double mzs,
                          double mhs, double mts, double GF,
                          int ExpDepth = -1);
\end{lstlisting}
defined in the header file \texttt{ff/ggzh/ggzhFF.h}, where \verb|GF| denotes $G_F$ and \texttt{ExpDepth} defines again the number of expansion terms in the $1/m_t^2$ expansion but the default value is set to \texttt{ExpDepth = -1}, which corresponds to the case where at each loop-order all possible expansion terms are taken into account. Internally this function computes the squared amplitude by performing the summation over the helicity amplitudes, which are calculated with the function
\begin{lstlisting}
vector<complex<double>> ggZHhels(vector<complex<double>> FF,
                                 double s, double t, double mzs,
                                 double mhs);
\end{lstlisting}
where the input vector \texttt{FF} is assumed to contain the form factors in the ordering of equation \eqref{eq:ffs} and returns the helicity amplitudes $\mathcal{A}_{+++}$, $\mathcal{A}_{++-}$, $\mathcal{A}_{++0}$, $\mathcal{A}_{+-+}$, $\mathcal{A}_{+--}$, $\mathcal{A}_{+-0}$. The polarization vectors of the gluons and the $Z$ boson are chosen as in Ref. \cite{Davies:2025out}. An example for the evaluation of the squared amplitude is given in the file \texttt{examples/ggzh-nnlo/Asq-lme.cpp}, which in combination with the python scripts in the same directory reproduces Fig. \ref{fig::matsq_pt10_3l} and \ref{fig::matsq_pt10_sum}.

\end{appendix}


\bibliographystyle{jhep}
\bibliography{inspire.bib,extra.bib}


\end{document}